\documentclass[twocolumn]{aastex631}
\usepackage[utf8]{inputenc}
\usepackage{graphicx}
\usepackage{fancyhdr}

\shorttitle{The \textit{Fermi}-LAT Light Curve Repository}
\shortauthors{The \textit{Fermi}-LAT Collaboration}
\graphicspath{{./}{}}

\begin{document}

\title{The \textit{Fermi}-LAT Light Curve Repository}


\author[0000-0002-6803-3605]{S.~Abdollahi}
\affiliation{IRAP, Universit\'e de Toulouse, CNRS, UPS, CNES, F-31028 Toulouse, France}
\author[0000-0002-6584-1703]{M.~Ajello}
\affiliation{Department of Physics and Astronomy, Clemson University, Kinard Lab of Physics, Clemson, SC 29634-0978, USA}
\author[0000-0002-9785-7726]{L.~Baldini}
\affiliation{Universit\`a di Pisa and Istituto Nazionale di Fisica Nucleare, Sezione di Pisa I-56127 Pisa, Italy}
\author[0000-0002-8784-2977]{J.~Ballet}
\affiliation{Universit\'e Paris-Saclay, Universit\'e Paris Cit\'e, CEA, CNRS, AIM, F-91191 Gif-sur-Yvette Cedex, France}
\author[0000-0002-6954-8862]{D.~Bastieri}
\affiliation{Istituto Nazionale di Fisica Nucleare, Sezione di Padova, I-35131 Padova, Italy}
\affiliation{Dipartimento di Fisica e Astronomia ``G. Galilei'', Universit\`a di Padova, Via F. Marzolo, 8, I-35131 Padova, Italy}
\affiliation{Center for Space Studies and Activities ``G. Colombo", University of Padova, Via Venezia 15, I-35131 Padova, Italy}
\author[0000-0002-6729-9022]{J.~Becerra~Gonzalez}
\affiliation{Instituto de Astrof\'isica de Canarias, Observatorio del Teide, C/Via Lactea, s/n, E-38205 La Laguna, Tenerife, Spain}
\author[0000-0002-2469-7063]{R.~Bellazzini}
\affiliation{Istituto Nazionale di Fisica Nucleare, Sezione di Pisa, I-56127 Pisa, Italy}
\author[0000-0001-8008-2920]{A.~Berretta}
\affiliation{Dipartimento di Fisica, Universit\`a degli Studi di Perugia, I-06123 Perugia, Italy}
\author[0000-0001-9935-8106]{E.~Bissaldi}
\affiliation{Dipartimento di Fisica ``M. Merlin" dell'Universit\`a e del Politecnico di Bari, via Amendola 173, I-70126 Bari, Italy}
\affiliation{Istituto Nazionale di Fisica Nucleare, Sezione di Bari, I-70126 Bari, Italy}
\author[0000-0002-4264-1215]{R.~Bonino}
\affiliation{Istituto Nazionale di Fisica Nucleare, Sezione di Torino, I-10125 Torino, Italy}
\affiliation{Dipartimento di Fisica, Universit\`a degli Studi di Torino, I-10125 Torino, Italy}
% ...

\begin{abstract}
The {\sl Fermi} Large Area Telescope (LAT) light curve repository (LCR) is a publicly available, continually updated library of gamma-ray light curves of variable {\sl Fermi}-LAT sources generated over multiple timescales. The {\sl Fermi}-LAT LCR aims to provide publication-quality light curves binned on timescales of 3 days, 7 days, and 30 days for 1525 sources deemed variable in the source catalog of the first 10 years of {\sl Fermi}-LAT observations. The repository consists of light curves generated through full likelihood analyses that model the sources and the surrounding region, providing fluxes and photon indices for each time bin.  The LCR is intended as a resource for the time-domain and multi-messenger communities by allowing users to quickly search LAT data to identify correlated variability and flaring emission episodes from gamma-ray sources.  We describe the sample selection and analysis employed by the LCR and provide an overview of the associated data access portal.
\end{abstract}


\keywords{{\sl Fermi}-LAT, gamma-rays, time domain, active galaxies}

\section{Introduction} \label{sec:intro}

Study of variability of astronomical objects has led to many discoveries in modern astronomy. Identified as one of the central themes of the 2020 Decadal Review, the increasing realization of time-domain and multi-messenger astronomy is now opening an entirely new window on the Universe. A high duty cycle and long-term monitoring of the gamma-ray sky has made the {\sl Fermi} Large Area Telescope \citep[LAT;][]{theLAT} a pivotal tool in the study of time-domain and multi-messenger astronomy. More than a decade of LAT observations has provided the identification and regular monitoring of thousands of transient, variable, and steady-state sources \citep{2022arXiv220111184F}. The long-term monitoring of the gamma-ray sky by the LAT played a crucial role in the first association of a high-energy neutrino detected by the IceCube neutrino observatory and the flaring blazar TXS 0506+056 \citep{IceCube:2018dnn, IceCube:2018cha}. 

The TXS 0506+056 association provided the first tantalizing clue to the origin of the high-energy cosmic neutrino flux detected by the IceCube neutrino observatory \citep{Astronuflux}. Though blazars have long been suggested as a possible source of extragalactic high-energy neutrinos, constraints on neutrino emission from bright gamma-ray blazars have largely disfavored steady-state active galactic nuclei (AGN) as the primary source of the observed neutrino flux \citep{2017ApJ...835...45A}. The TXS 0506+056 association, however, has shown that at least some of the astrophysical neutrinos detected by IceCube could be attributed to high fluence AGN that undergo periods of intense flaring activity on top of a much lower quiescent gamma-ray emission that may be undetectable to the LAT.

The long-term monitoring provided by the LAT has also been indispensable to multi-wavelength campaigns that aim to study long-term correlated variability in AGN \citep[e.g.,][]{Kramarenko2022}. Observations of radio and optical flares have been used to examine the location of the gamma-ray emitting region, as well as the particle acceleration and radiation processes in the relativistic jets in these sources. For example, the long-term correlated variability between optical and gamma-ray flares has allowed for the detection of a systematic lag in the optical-to-gamma-ray flares in 3C 279 \citep{2012ApJ...754..114H,2015ApJ...807...79H}, the evidence of quasi-periodic variations in the BL Lac object PG 1553+113 \citep{2015ApJ...813L..41A}, and constraints on the rate of orphan optical flares with no gamma-ray counterpart \citep{2019ApJ...880...32L}. Such studies of correlated multi-wavelength variability are a crucial diagnostic of the physics of relativistic jets. 
 
Despite the importance of long-term variability studies of LAT sources, very few existing resources enable the community to easily access the whole-mission data to perform correlative analyses. Resources like the Fermi All-Sky Variability Analysis \citep[FAVA;][]{Abdollahi2017} and the monthly aperture photometry light curves of 3FGL sources \footnote{\url{https://fermi.gsfc.nasa.gov/ssc/data/access/lat/4yr_catalog/ap_lcs.php}. 
} allow users to quickly examine a source for relative flux increases, but do not provide flux calibrated characterization of a source.  Doing so requires a full likelihood analysis of the region that takes into account the flux variations of all nearby sources, and generating a high cadence (e.g., daily) light curve using a full likelihood treatment over the entire duration of the mission can be very time consuming. 

To address this need, we developed the {\sl Fermi}-LAT Light Curve Repository \citep[LCR; ][]{LCR2021}\footnote{\url{https://fermi.gsfc.nasa.gov/ssc/data/access/lat/LightCurveRepository/}}, consisting of a public database of light curves for variable {\sl Fermi}-LAT sources on a variety of timescales. The repository provides publication-quality light curves on timescales of 3 days, 7 days (weekly), and 30 days (monthly) for 1525 sources deemed variable in the 4FGL-DR2 catalog \citep{4fgl}. The repository consists of light curves generated through full likelihood analyses of the sources and surrounding regions, providing flux and spectral index measurements for each time bin. Hosted at NASA's Fermi Science Support Center (FSSC), the LCR provides users with on-demand access to this light curve data, which is continually updated as new data becomes available. The repository is a new resource to the time-domain and multi-messenger communities for associating and monitoring LAT sources, in particular facilitating identification of time intervals with high fluence or flux. 

This paper is organized as follows. In Sec.~\ref{sec:sample} we describe the gamma-ray sources included in the LCR. Sec.~\ref{sec:analysis} is devoted to the description and discussion of the automated data analysis employed by the repository. We discuss analysis caveats and data usage best practices in Sec.~\ref{sec:caveats}. We briefly summarize and conclude in Sec.~\ref{sec:conclusion}. We provide a \textit{Quick Guide} in Appendix~\ref{sec:guide} with the aim to help familiarize users with the LCR data portal.

\section{The Source Sample}
\label{sec:sample}

\begin{figure}[t]
\begin{center}
\includegraphics[width=\linewidth,angle=0]{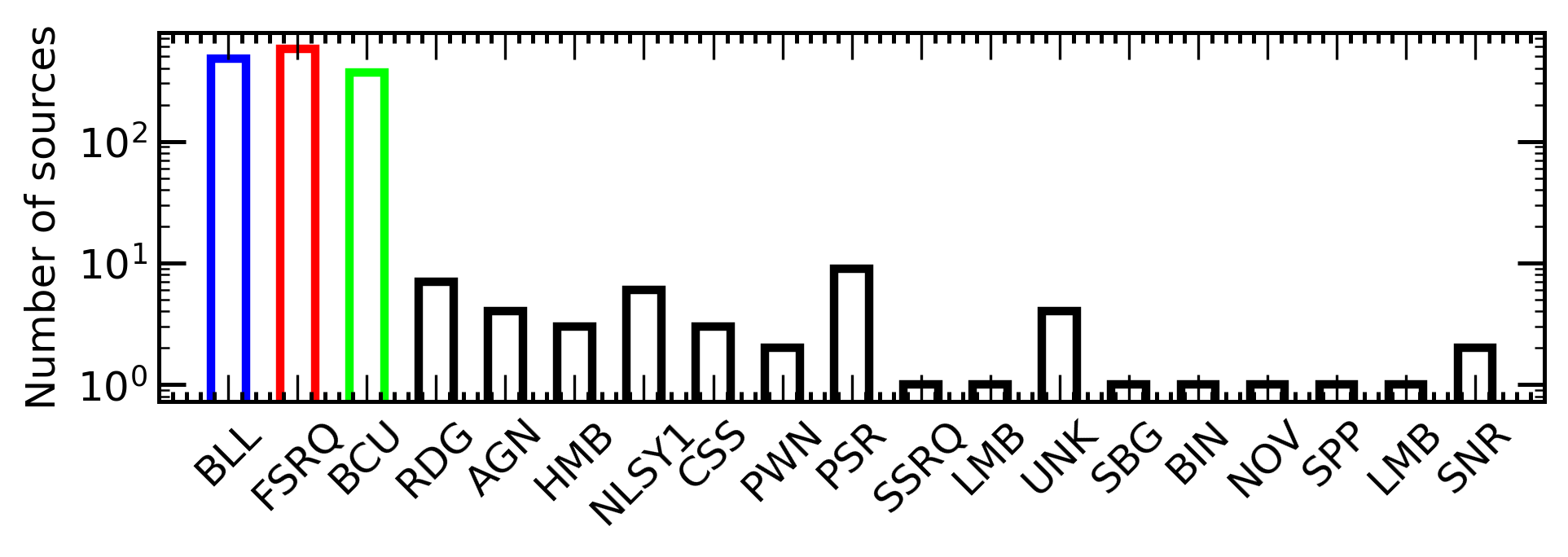}\\
\includegraphics[width=\linewidth,angle=0]{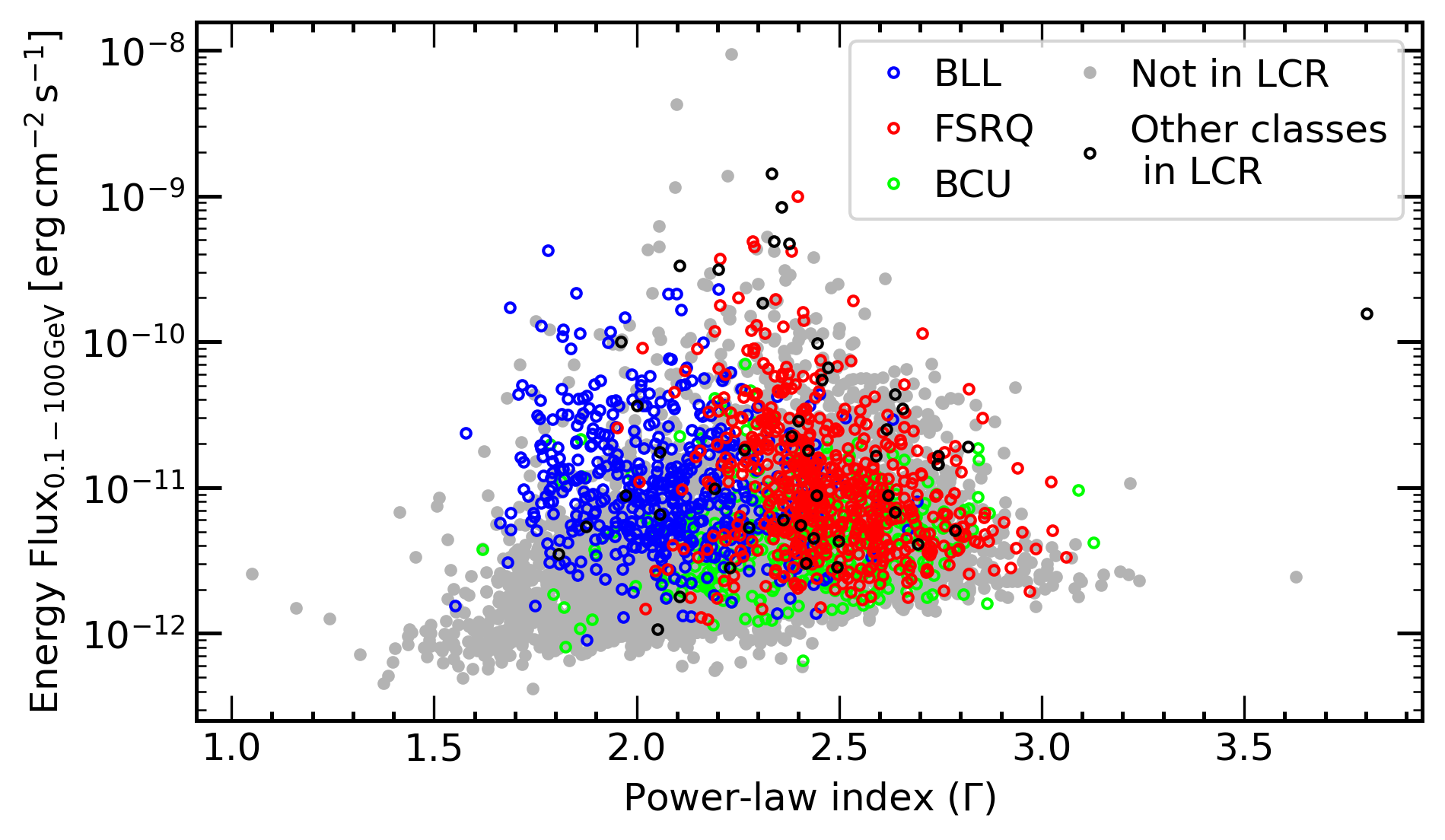}\\
\includegraphics[width=\linewidth,angle=0]{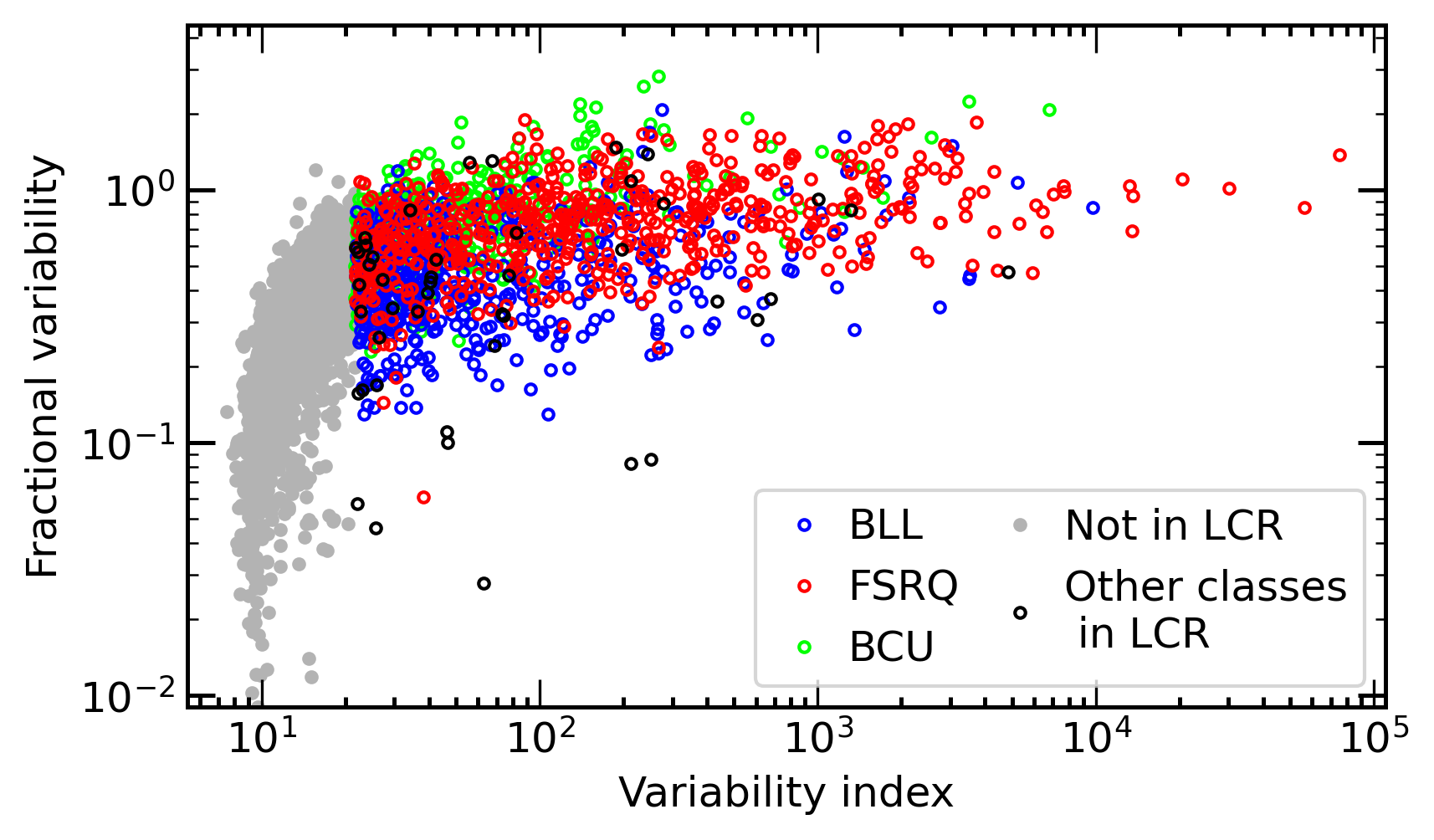}
\caption{Top: Histogram of the source types included in the LCR population, defined as having a variability index above 21.67, corresponds to $<$1\% chance of steady emission. Center and bottom: The population of sources included in the LCR (empty markers). In grey we report the 4FGL-DR2  population not included in the LCR as reference. The energy range of integration is 0.1$-$100 GeV.
\label{fig:demog}}
\end{center}
\end{figure}

Motivated by the science described above, the LCR focuses on the sources in the 4FGL-DR2 catalog that have variability indices greater than 21.67, where the variability index can be thought of as a proxy for the average fractional variability $\delta F/F$, with $\delta F$ measured on timescales of 1 year.  As defined in the 4FGL-DR2 catalog, which is based on 10 years of survey data, sources with such a variability index over 10 years are estimated to have a less than 1\% chance of being steady. The resulting sample consists of 1525 sources, or roughly 26\% of the 4FGL-DR2 catalog. The vast majority of these sources is blazars, further classified as flat spectrum radio quasars (FSRQ), BL Lacs (BLL), and blazar candidates of unknown type (BCU), making up roughly 38\%, 31\%, and 24\% of the repository sample, respectively, or 77\%, 36\% and 26\% of their respective 4FGL-DR2 class. This is consistent with the fact that the LAT class with the largest number of variable sources are FSRQs. As illustrated in the top panel of Fig.~\ref{fig:demog}, we also count 7 radio galaxies (RDG), 4 AGN, 3 high-mass binaries (HMB), 6 narrow-line Seyfert 1 galaxies (NLSY1), 3 compact steep spectrum radio sources (CSS), 2 pulsar wind nebulae (PWN), 9 pulsars (PSR), 1 steep spectrum radio quasar (SSRQ), 1 low-mass binary (LMB), 4 associations to unclassified sources (UNK), 1 starburst galaxy (SBG), 1 binary (BIN), 1 nova (NOV), 1 supernova remnant/pulsar wind nebula (SPP), and 2 supernova remnants (SNR). In the middle and bottom panels of the same figure, we show plots of the spectral shape parameters (power-law index and total energy flux) and variability parameters (variability index and fractional variability from DR2) for the sources included in the LCR (colored points) overplotted on the full 4FGL population (grey points). The former shows that most of the sources not included are fainter with harder spectra, i.e., mainly BL Lac objects. This is because, as can be seen in the bottom panel of the same figure, BL Lac objects are intrinsically less variable than FSRQs in the {\sl Fermi}-LAT energy range. 

\section{Automated Data Analysis} \label{sec:analysis}

Generating 3-day, 7-day, and 30-day light curves for each of these sources for over 13 years of data requires the analysis of over 3.7 million individual time bins. The LCR analysis pipeline runs on a cluster hosted at SLAC National Accelerator Laboratory. Generating the light curves over the entire mission to date for the included sources requires approximately 3 months ($>$400 core years).

In the following subsections we present the details of the analysis and computational strategies used to generate the LCR data products.

\subsection{Analysis Technique \& Tools}

The characterization of LAT sources is typically performed using a maximum likelihood analysis \citep{2009ApJS..183...46A}, in which the parameters of a model describing the point sources and diffuse isotropic gamma-ray emission in a given region of the sky are jointly optimized to best describe the observed photon distribution. The light curves of the LCR are obtained by performing an unbinned likelihood analysis, in which the full spatial and spectral information of each photon is used in the maximum likelihood optimization.

The LCR analysis is performed with the standard LAT {\tt Fermitools}\footnote{\url{https://fermi.gsfc.nasa.gov/ssc/data/analysis/software/}} (version {\tt 1.0.5}) using the \texttt{P8R3\_SOURCE\_V2} instrument response functions on \texttt{P8R3\_SOURCE} class \citep{2013arXiv1303.3514A, 2018arXiv181011394B} photons selected over the energy range covering 100 MeV--100 GeV. Note that the energy dispersion is neglected in the unbinned analysis mode. This is not expected to impact the quality of the analysis. However, repeating the analysis with a binned likelihood approach would return mostly insignificant differences in the results. For each source and time bin, photons are selected from a circular region of interest (ROI) of radius 12$^{\circ}$ centered on the location of the target source. The ROIs are analyzed separately. By contrast, for the standard FGL catalogs, fluxes and spectral parameters of multiple sources are extracted from the same ROI. The size of the ROI is conservatively chosen to be as large as the 95\% containment radius of the LAT energy-dependent point-spread function (PSF) at 100 MeV. Additional data selection cuts are imposed to exclude photons associated with regions and periods of known solar flares and gamma-ray bursts. A zenith angle limit of 90$^{\circ}$ strongly reduces contamination from gamma rays produced through interactions of cosmic rays with Earth’s atmosphere (Earth limb).

\begin{figure}[t]
\begin{center}
\includegraphics[width=\linewidth,angle=0]{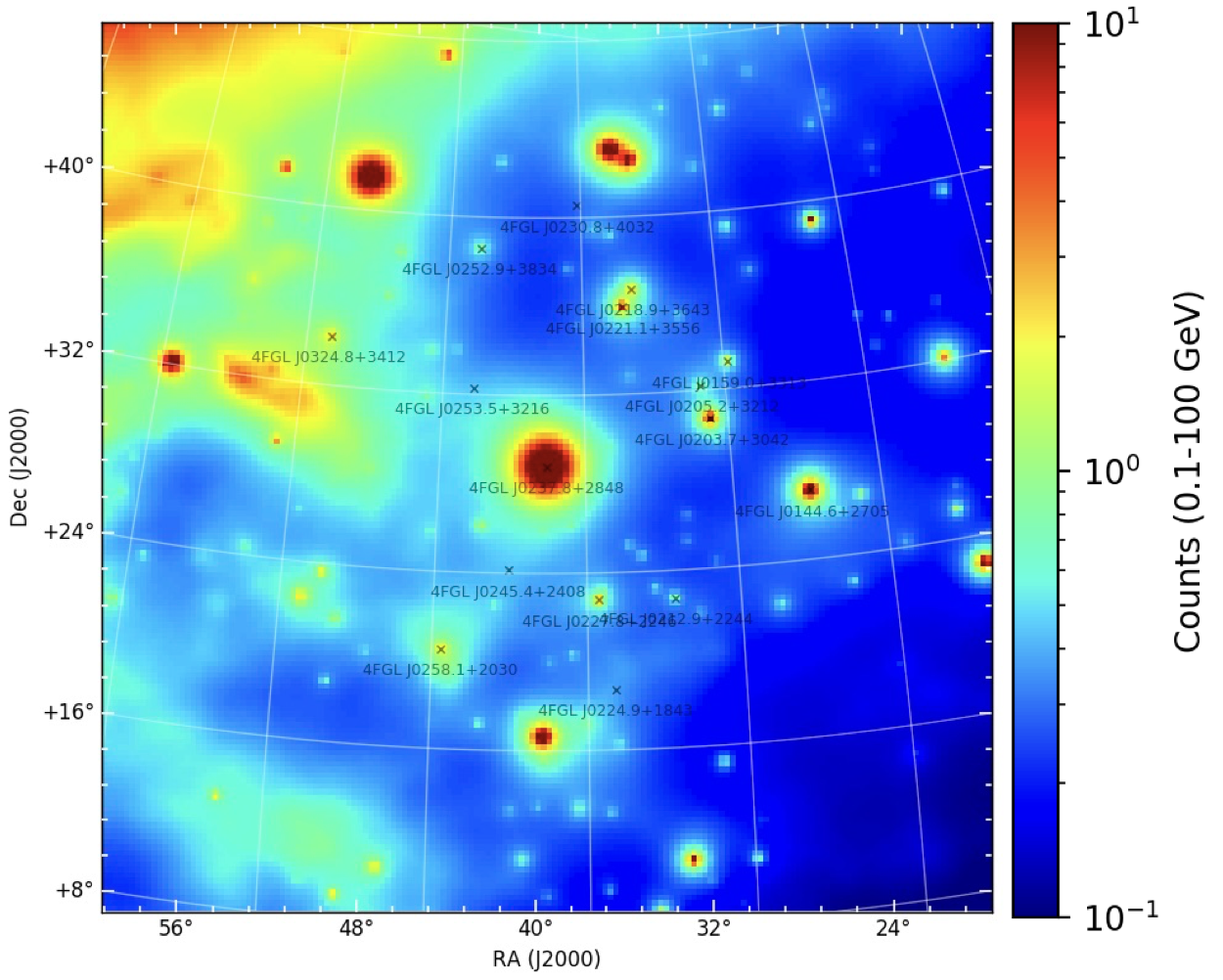}
\caption{A model map, with 0.1 degree resolution, for a single weekly time bin of the region surrounding FSRQ 4C~+28.07. This source contains 15 other variable sources within a 12$^{\circ}$ radius, highlighting the need to model all variable sources within the ROI. 
\label{fig:map}}
\end{center}
\end{figure}

\begin{figure*}[t]
\begin{center}
\includegraphics[width=\textwidth,angle=0]{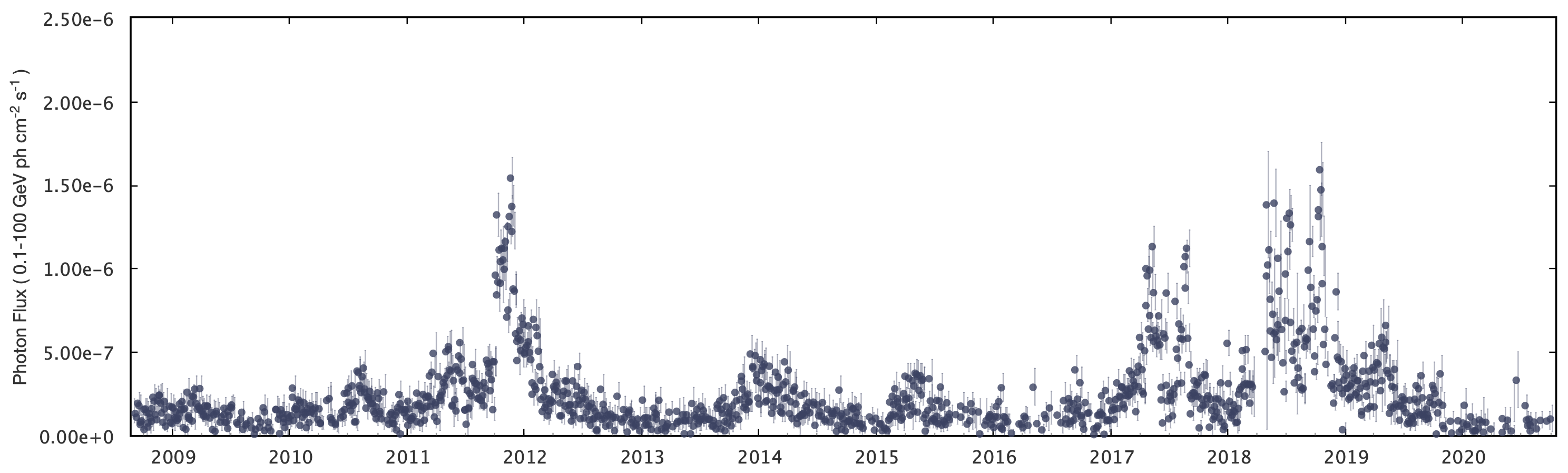}
\caption{An example 3-day light curve as can be found in the LCR. It spans more than 11 years of LAT data and refers to the source at the center of Figure \ref{fig:map}, FSRQ 4C~+28.07. This light curve is obtained by specifying the analysis option that the spectral index of the source is free to vary. The data gap in 2018 is due to the temporary shut down of the instrument because of a solar panel anomaly.
\label{fig:lc}}
\end{center}
\end{figure*}

Following the 4FGL-DR2 catalog, the LCR sources can have one of three different spectral types:
\begin{itemize} \itemsep -0pt
    \item Power-law (PL):
    \begin{equation}
        dN/dE=N_0(E/E_0)^{-\Gamma}
    \end{equation}
    \item log-parabola (LP): 
    \begin{equation}
        dN/dE=N_0(E/E_b)^{-(\alpha+\beta \log(E/E_b))}
    \end{equation}
    \item and subexponentially cutoff power-law (PLEC): 
    \begin{equation}
         dN/dE=N_0(E/E_0)^{-\gamma_1}e^{-aE^{\gamma_2}}
    \end{equation}
\end{itemize}
A majority of LCR sources are best described by a PL or LP spectral shape. The PLEC spectral shape best represents 11 additional sources, comprising the 9 pulsars in the LCR sample and the two brightest FSRQs (CTA 102 and 3C 454.3).

The normalization of the source spectrum in the model is left free to vary, while the spectral shape parameters are initially fixed to their 4FGL-DR2 catalog values. The model also includes all gamma-ray sources in the 4FGL-DR2 catalog within a radius of 30$^{\circ}$ from the ROI center. The normalization of each variable source in the ROI is also left free to vary in the model, with spectral shape parameters fixed to their catalog values. In addition to the point sources, Galactic and isotropic background components are included in the model. The Galactic component, \texttt{gll\_iem\_v07.fits}\footnote{\url{https://fermi.gsfc.nasa.gov/ssc/data/access/lat/BackgroundModels.html}}, is a spatial and spectral template that accounts for interstellar diffuse gamma-ray emission from the Milky Way. The isotropic component, \texttt{iso\_P8R3\_SOURCE\_V3\_v1}, provides a spectral template to model the remaining isotropic events, including contributions from the residual charged-particle background and the isotropic celestial gamma-ray emission. The normalizations of both the Galactic and isotropic components are allowed to vary during the fit. The free parameters of the model are varied to maximize the likelihood of observing the data given the model. An iterative fitting strategy, which varies the required fit tolerance\footnote{The fit tolerance is the relative convergence tolerance that is specified in the {\tt gtlike} routine and passed to the optimization algorithm used to maximize the log likelihood function. The fit tolerance can be loosened to help achieve fit convergence. On the other hand, fits with tighter fit tolerances exhibit overall lower fractional errors on the resulting flux estimations. Therefore an iterative approach has been adopted for the LCR.} over three steps (1, $10^{-4}$ and $10^{-8}$), is employed to minimize the number of time bins in which the likelihood fit does not successfully converge. Once fit convergence is achieved with the tightest tolerance, a second round of fitting is performed in which a spectral shape parameter of the target source is allowed to vary, namely, photon index ($\Gamma$) for PL, $\alpha$ for LP, and $\gamma_1$ for the PLEC model. All other parameters remain fixed, i.e., $\beta$ for LP, and $a$ and $\gamma_2$ for the PLEC model.

Figure \ref{fig:map} shows a model map for a single time bin of the region surrounding FSRQ 4C +28.07 (4FGL J0237.8+2848), which contains 15 other variable sources within a 12$^{\circ}$ radius. Figure \ref{fig:lc} shows the resulting 3-day light curve spanning more than 11 years of LAT data for 4FGL J0237.8+2848. 

A likelihood ratio test \citep{Neyman1928} is used to quantify the significance of the target source above the background; specifically the test statistic \citep{1996ApJ...461..396M}: 
\begin{equation}
{\rm TS} = -2~{\rm log}(L_0/L). \label{eq:ts}
\end{equation}
The TS compares the maximum value of the likelihood function $L_0$  evaluated for the parameter values that maximize the likelihood under a background-only null hypothesis (i.e., a model that does not include a target source), with $L$, the likelihood function evaluated at the best-fit model parameters when including the target source. In the null hypothesis, TS is distributed approximately as $\chi^2$ \citep{wilks1938}, and the analysis rejects the null hypothesis when the test statistic is greater than ${\rm TS}\geq4$, which is roughly equivalent to a 2$\,\sigma$ rejection criterion for a single degree of freedom\footnote{The null hypothesis is tested against the presence of one source at a known position (at the center of the ROI)}. Using this threshold value of TS as the detection criterion, the LCR employs a Bayesian profile likelihood method\footnote{A description of the Bayesian profile likelihood method employed by the LCR can be found at \url{https://fermi.gsfc.nasa.gov/ssc/data/p7rep/analysis/scitools/python_tutorial.html}.} to calculate the 95\% confidence level upper limits for any interval that yields a ${\rm TS}\leq4$, and also extracting a flux estimation of the target source for any interval with a ${\rm TS}\geq1$. The resulting best-fit values (or upper limits) for photon flux (cm$^{-2}$~s$^{-1}$), energy flux
(GeV cm$^{-2}$~s$^{-1}$), and associated spectral shape are saved to the LCR database, for both the fixed and free photon index analyses. This procedure ultimately allows for a user-selectable detection threshold and spectral fit method, as the flux estimates and upper limits between $1\le{\rm TS}\le4$ are both recorded in the LCR database, as well as the results from both the fixed and free spectral analyses.

\begin{figure}[t]
\begin{center}
\includegraphics[width=\linewidth,angle=0]{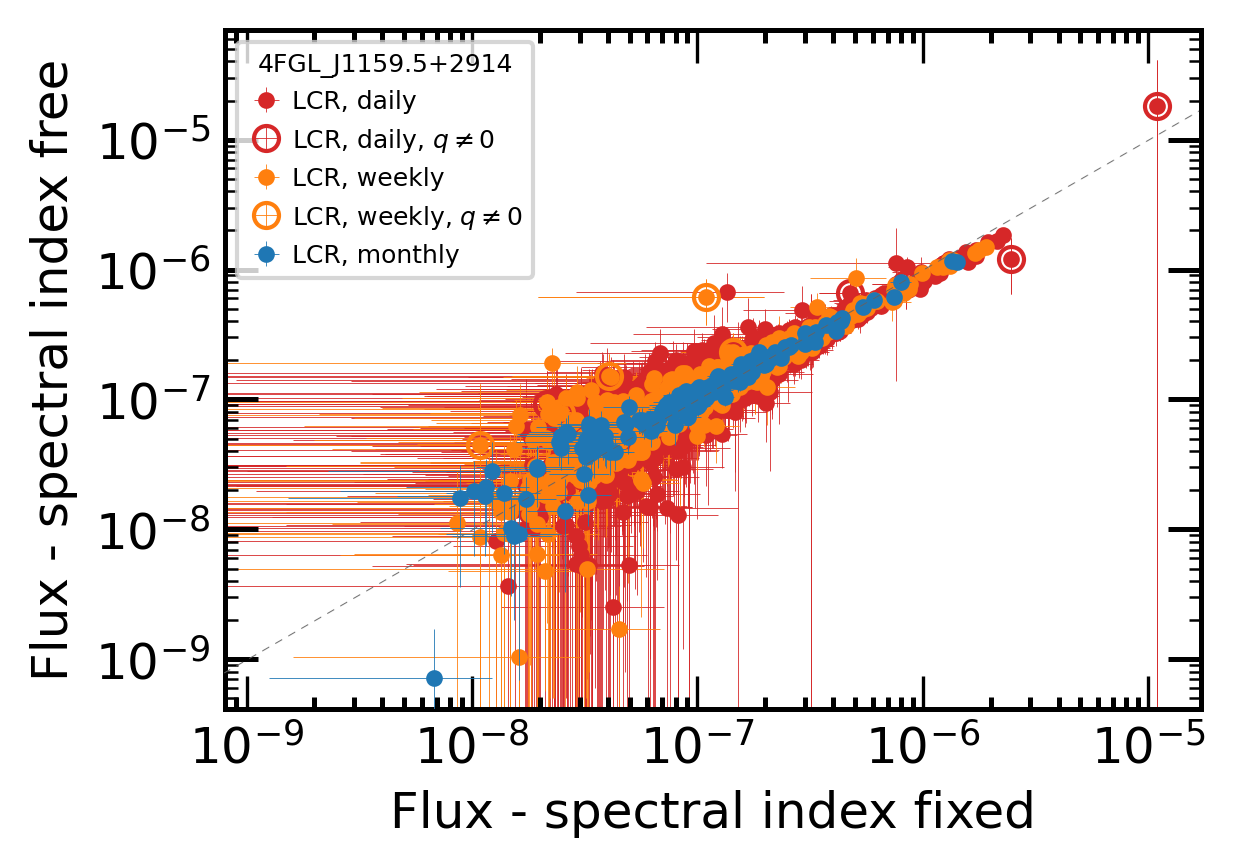}
\caption{Comparison of the fluxes [ph/cm$^{2}$/s] in time bins obtained with the fitting pipeline keeping the spectral index fixed versus letting the spectral index be free. Empty circles indicate results from analyses that did not converge, i.e., \texttt{MINUIT Return Code} $q\neq 0$. \texttt{MINUIT} \citep{minuit} is the optimizer adopted in the likelihood analysis. The plot  shows the fluxes measured for the quasar Ton~599. In this case, the fit for the highest-flux point (with TS$\sim 4$) converged when the spectral index was set free, but did not converge when the spectral index was fixed.
}
\label{fig:plindex}
\end{center}
\end{figure}

In Fig.~\ref{fig:plindex} we compare, for the case of the bright quasar Ton 599, the best-fit flux values found using a fixed spectral index vs. leaving the spectral index  free to vary. The colors in the plot refer to the three cadences analysed. The results are generally stable, with greater uncertainties for lower fluxes for higher cadences due to lower statistics in the ROI in each time bin. 

\subsection{Computational Strategy}
For analyses of a relatively small number (a few thousands) of photons, an unbinned likelihood analysis can be performed rapidly (a few minutes), but as the number of events increases, the time to perform the analysis can become prohibitive. This limitation becomes increasingly burdensome when the need arises to perform a source characterization over a large number of time bins. A binned likelihood analysis could alleviate this issue. However, information is lost when binning data. The LCR tackles the computational overhead by parallelizing the process of performing a full unbinned likelihood analysis. In order to produce a high cadence light curve over the entire lifetime of the mission in a reasonable amount of time, the LCR distributes the analyses of the light curve bins to separate nodes in a computer cluster hosted at the SLAC National Accelerator Laboratory and utilizing the IBM Spectrum LSF workload management platform. The parallelization allows for thousands of time bins to be analyzed simultaneously, with the net effect of drastically reducing the time to generate light curves over the entire duration of the mission.
\begin{figure*}[htbp]
\begin{center}
\includegraphics[width=\textwidth,angle=0]{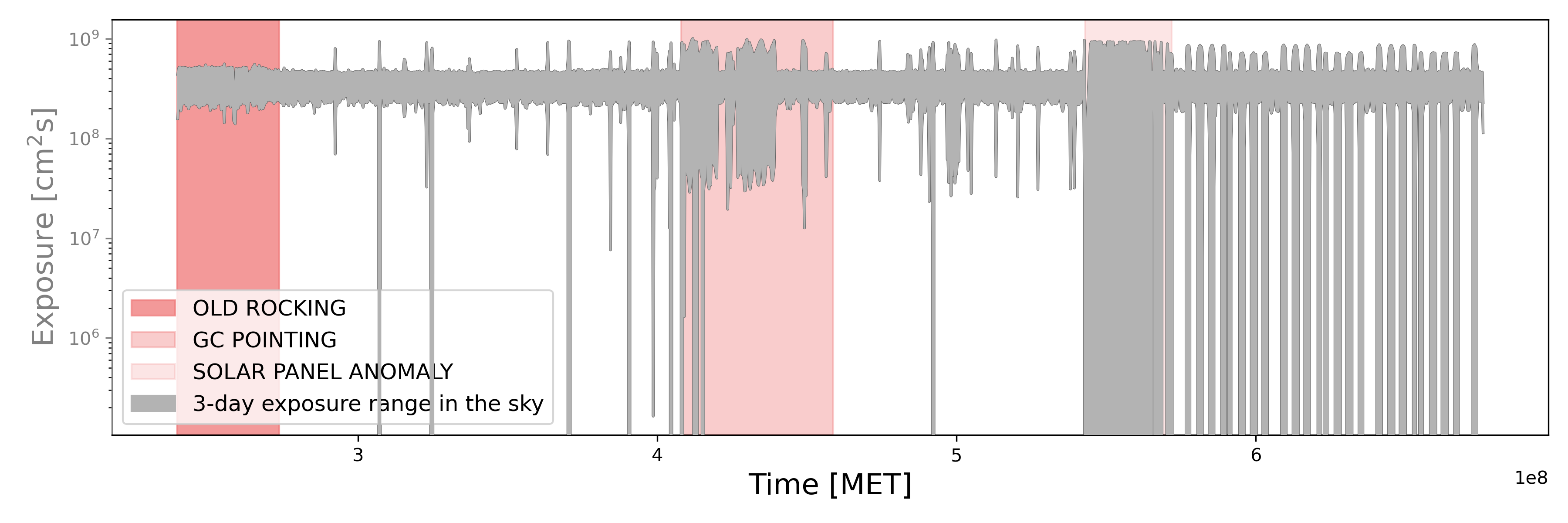}
\caption{The gray band marks the range of LAT on-sky exposures for each time bin of 3-day cadence. The pink bands, in order, mark time ranges of the original rocking angle mission profile$^{(\dag )}$, the Galactic center monitoring, and the time gap between the solar panel anomaly and the beginning of the new survey mode. Note that some time bins have a minimum exposure value of zero: this means that some part of the sky was not observed during that time interval. The exposure maps are computed at the central energy in the 0.1--100 GeV band. \\
$^{(\dag )}$ {\footnotesize Details on the Observatory sky-survey profiles can be found at \url{https://fermi.gsfc.nasa.gov/ssc/observations/types/allsky/}.}
\label{fig:exp_lc}}
\end{center}
\end{figure*} 

\subsection{Exposure analysis}
\label{subsec:exposure}
The {\it Fermi} spacecraft has been executing a sky-scanning strategy for the LAT for the great majority of the mission, generally reaching an almost uniform full-sky coverage daily, in fact, every two orbits (i.e., approximately every three hours). During the mission the scanning strategy has been interrupted occasionally for targeted observations. The longest-duration such observation was a modified observing strategy executed from 2013 December to 2015 July that favored the Galactic center region. In March 2018, the seizing of the drive motor for one of the solar panels forced the temporary shut down of the LAT and a redefinition of the survey mode\footnote{\url{https://fermi.gsfc.nasa.gov/ssc/observations/types/post_anomaly/}}. This caused the exposure to be limited to a part of the sky for some period of time. The effects of these periods of uneven exposure can be seen in the 3-day light curves. Therefore we quantify and discuss these effects in this section.

\begin{figure}[ht]
\begin{center}
\includegraphics[width=\linewidth,angle=0]{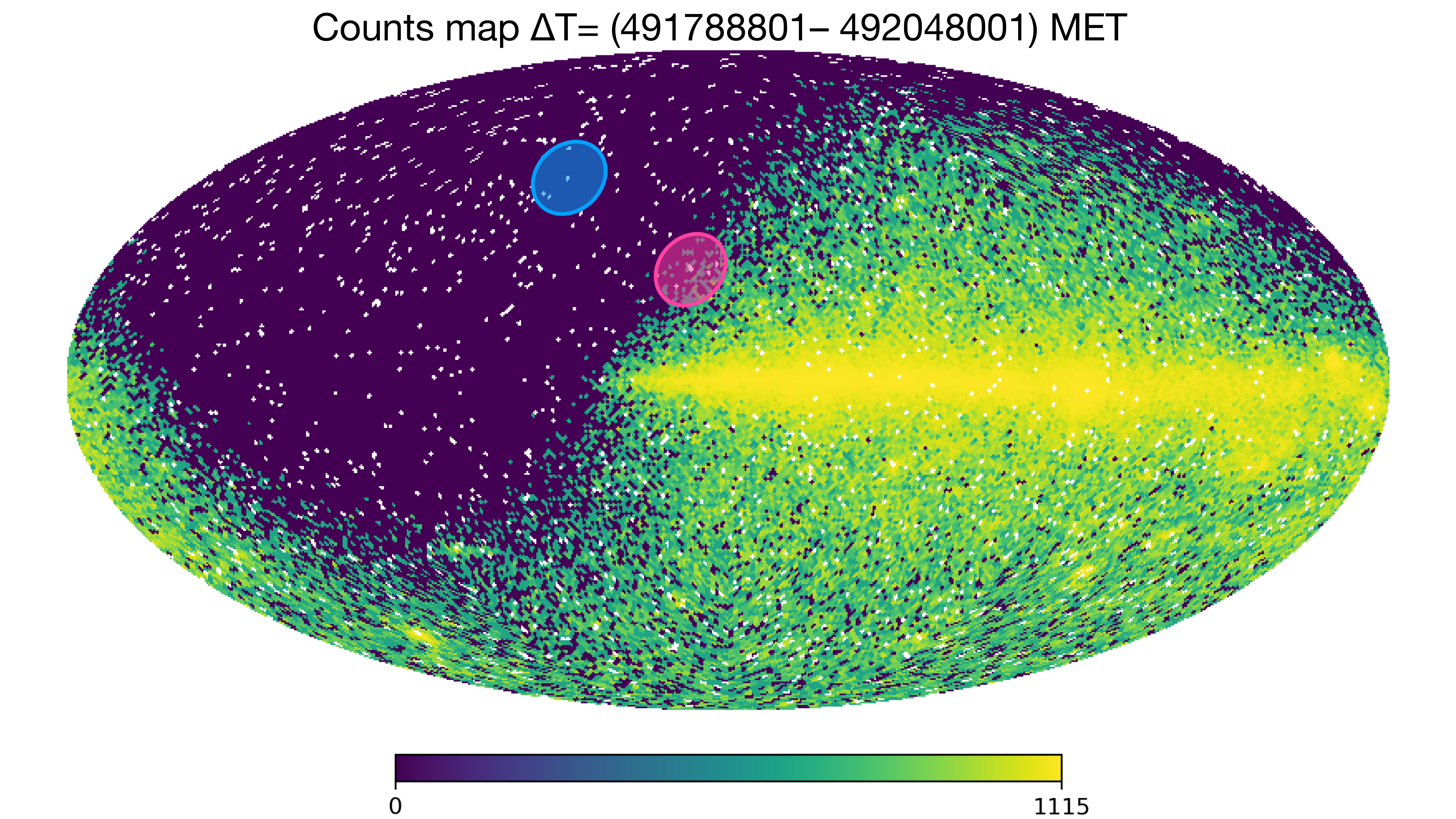}\\
\quad \\
\includegraphics[width=\linewidth,angle=0]{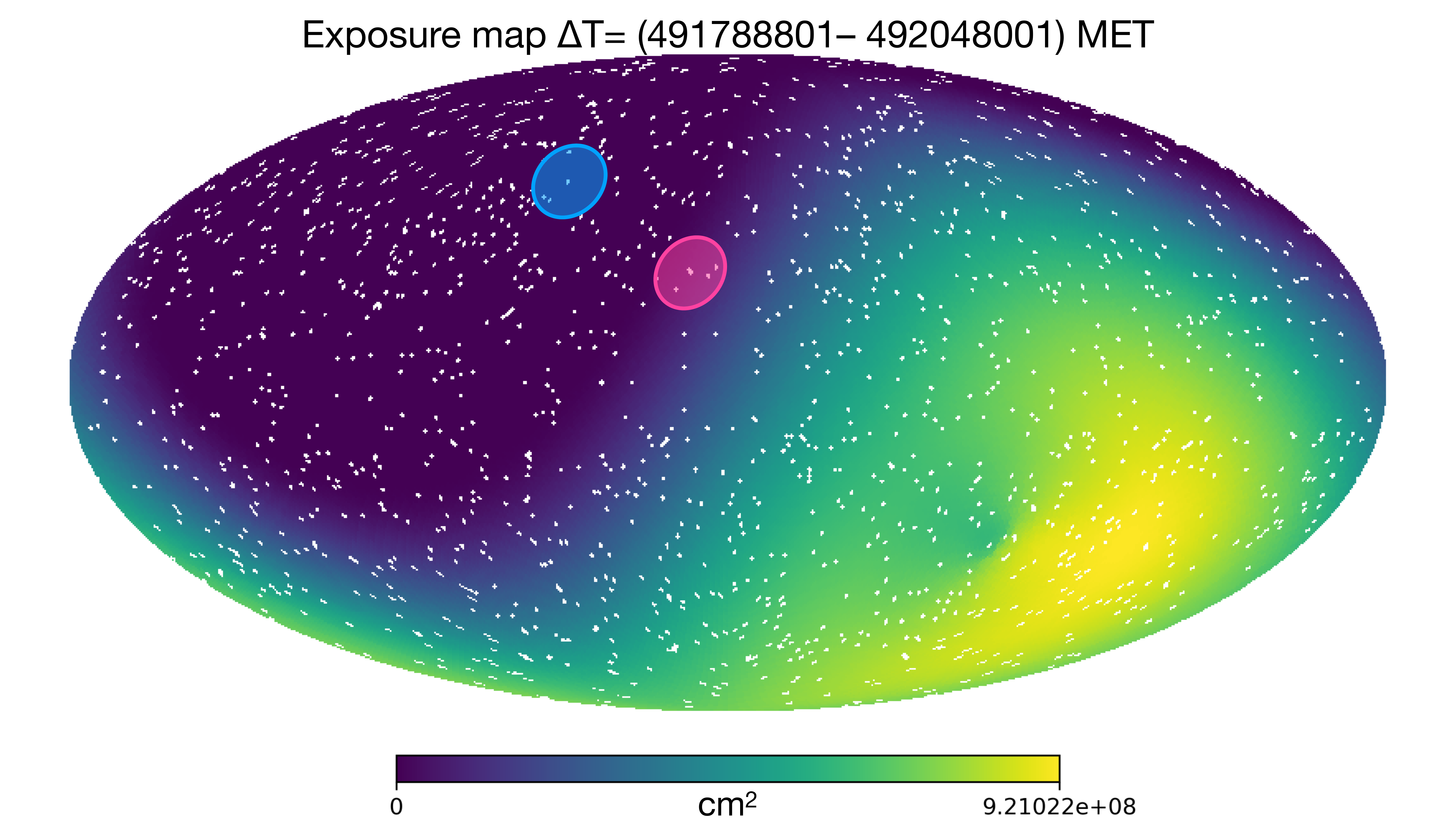}
\caption{Top: Counts map for a 3-day time bin with 0 minimum exposure; Bottom: Exposure map for the same time bin. The white pixels mark the positions of LCR sources. The light blue and magenta circles in the maps highlight the cases of two ROIs lying in a region of the sky with zero and non-flat exposures, respectively. These maps are Mollweide projections of the entire sky in Galactic coordinates, centered at the Galactic Center. Longitudes increase from right to left. The color scales in both maps are linear.}
\label{fig:maps}
\end{center}
\end{figure}

The number of photons in a given ROI is modulated by the exposure of the observation, which is given by the product of the LAT livetime (during the brief readout time of a photon or cosmic-ray interaction the LAT, the instrument is `dead' to triggering on other interactions) and the energy and angle-dependent effective area of the LAT. 
In general, analyses integrating over long observation times can have small fractional exposure variations across an ROI, but shorter-timescale analysis can be significantly impacted by low exposure for a particular region of the sky. 
In fact, since the beginning of the post-anomaly modification of the sky-scanning strategy in February 2019, a number of the 3-day and 7-day cadence time bins have no exposure in particular portions of the sky. This is due to the constraints on the direction of the zenith of spacecraft with respect to the Sun\footnote{Note that the `gaps' in the exposure for different time bins are not always in the same portion of the sky}. Fig.~\ref{fig:exp_lc} shows the range of values of the LAT exposure in the sky for each time bin of 3-day cadence. The maps in Fig.~\ref{fig:maps} illustrate the positions of all the sources analyzed in the LCR, overlaid on the counts map (top) and an exposure map (bottom) for a typical 3-day cadence time bin with a zero minimum exposure: some sources fall in the region of the sky with zero exposure, which naturally translates into zero events to analyze in the ROI.

In these cases, or in those with minimum exposure orders of magnitude lower than the maximum (shaded pink regions in the maps in Fig.~\ref{fig:maps}), the pipeline typically returns an upper limit on the source flux. However, in some cases it could still find a solution that maximizes the likelihood, but warns of a poor fit quality or an unreasonably low error value. For this reason, we recommend caution when working with the 3-day and 7-day cadence light curves for these low-exposure intervals, or exclude the affected time bins entirely from their analyses.

To illustrate the effect of low exposure on the photon counts, we generated binned all-sky counts and related exposure maps in the same time bins used for the LCR data analysis. 
We consider events in the 0.1--100 GeV energy range, and the maps are produced in HEALPix\footnote{\url{http://healpix.jpl.nasa.gov/} \citep{healpix}} format with the tool {\tt gtexpcube2}\footnote{The {\tt Fermitools}, as part of the maximum likelihood calculation, automatically account for the exposure, as described on the official mission web page at \url{https://fermi.gsfc.nasa.gov/ssc/data/analysis/documentation/Cicerone/Cicerone_Data_Exploration/livetime_and_exposure.html}. The module {\tt gtexpmap} is used to compute the exposure when performing unbinned analyses, as for the LCR analysis pipeline. In this section, we compute the exposure through the {\tt gtexpcube2} module, which is used when performing binned likelihood analyses. We stress that this is for purely illustrative purposes; the exposure maps computed in this section were not used in the LCR likelihood analyses.}, using the same analysis setup and IRFs as for the automated likelihood analysis. 
The HEALPix format allows us to easily extract the average exposure in every ROI, and the pixel resolution (HEALPix order 6) matches the resolution of the pre-computed livetime cubes provided for LAT data analyses.

\begin{figure}[ht]
\begin{center}
\includegraphics[width=\linewidth,angle=0]{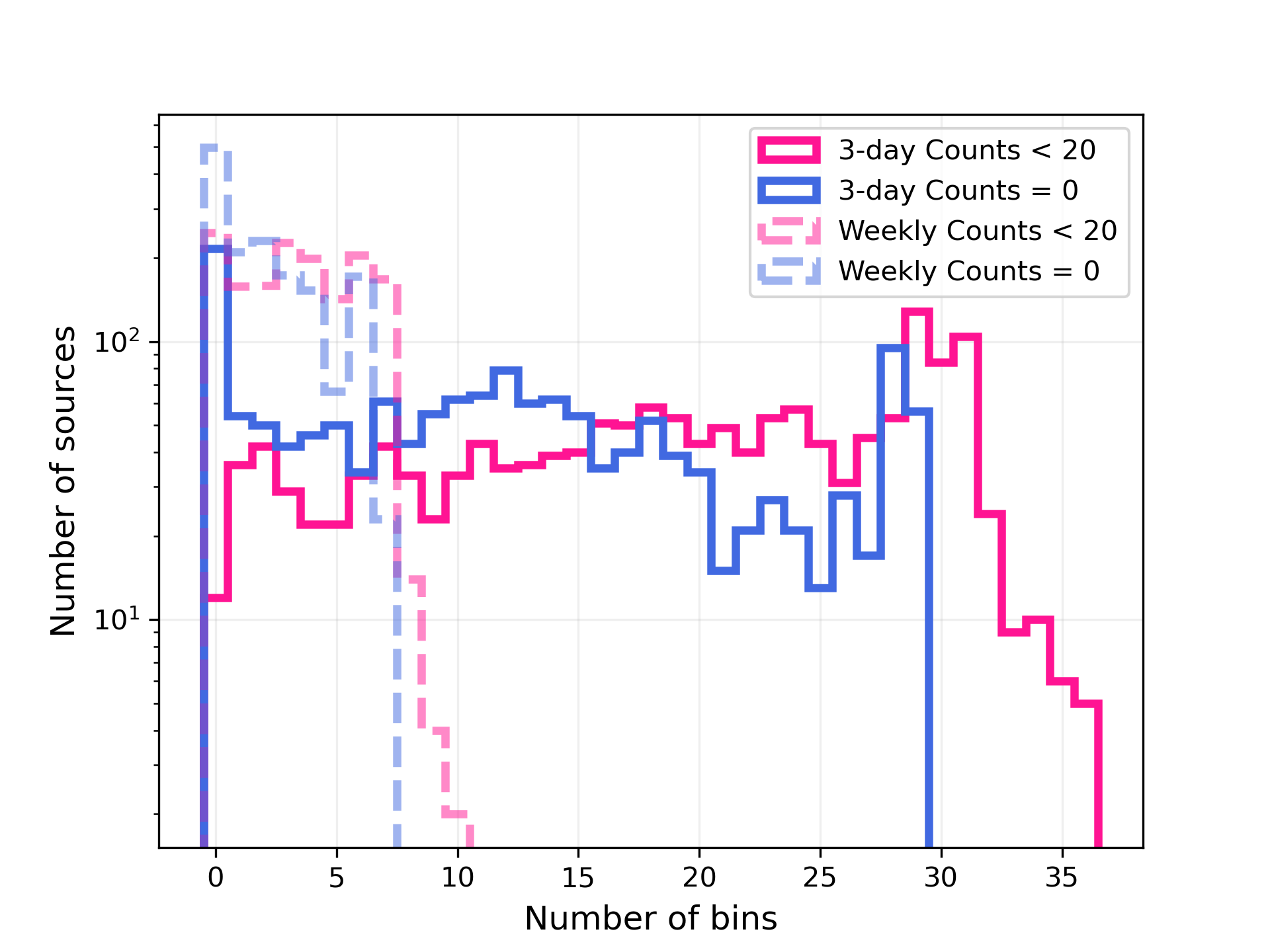}
\caption{In light blue, we show the distribution of LCR sources by number of time bins with zero photons within the ROI used for their analysis. In magenta, we show the distribution of LCR sources according the number of time bins with fewer than 20 photons within the ROI used for their analysis. Darker shades of the lines refer to the 3-day cadence, while lighter dashed lines refer to the weekly cadence. The low-exposure time bins represent the 0.1\% of the total time bins.
\label{fig:statBad}}
\end{center}
\end{figure}

For each source considered in the LCR and each time bin of 3-day and 7-day cadences, we extract the average exposure and the total photon counts in the ROI centered on the source between 0.1 and 100 GeV. In Fig.~\ref{fig:statBad} we illustrate the statistics of sources with a given number of time bins in the light curve that had no photons in the ROI or had fewer than 20 photons (considered as an arbitrary threshold for low-statistics analysis). Hundreds of sources have up to 30 time bins with zero events in the light curve and/or less than 20 events, while tens of sources have fewer than 20 photons in more than 50 time bins (note that the total number of time bins in a 3-day cadence light curve is more than 1680 as of now). The monthly cadence, due to the longer integration time, should not be affected by this issue; however, we still recommend that users check the exposure of each bin used in their analyses.
The exposures within the 12$^\circ$ radius ROIs for each source for the 3-day and 7-day cadences are provided in the LCR downloadable data.

\section{Usage caveats}
\label{sec:caveats}
In this section we provide a number of caveats to be mindful of when using the LCR data for scientific research. This list is also available and will be updated periodically on the LCR data portal. 

    \begin{figure*}[t]
    \begin{center}
    \includegraphics[width=0.33\textwidth,angle=0]{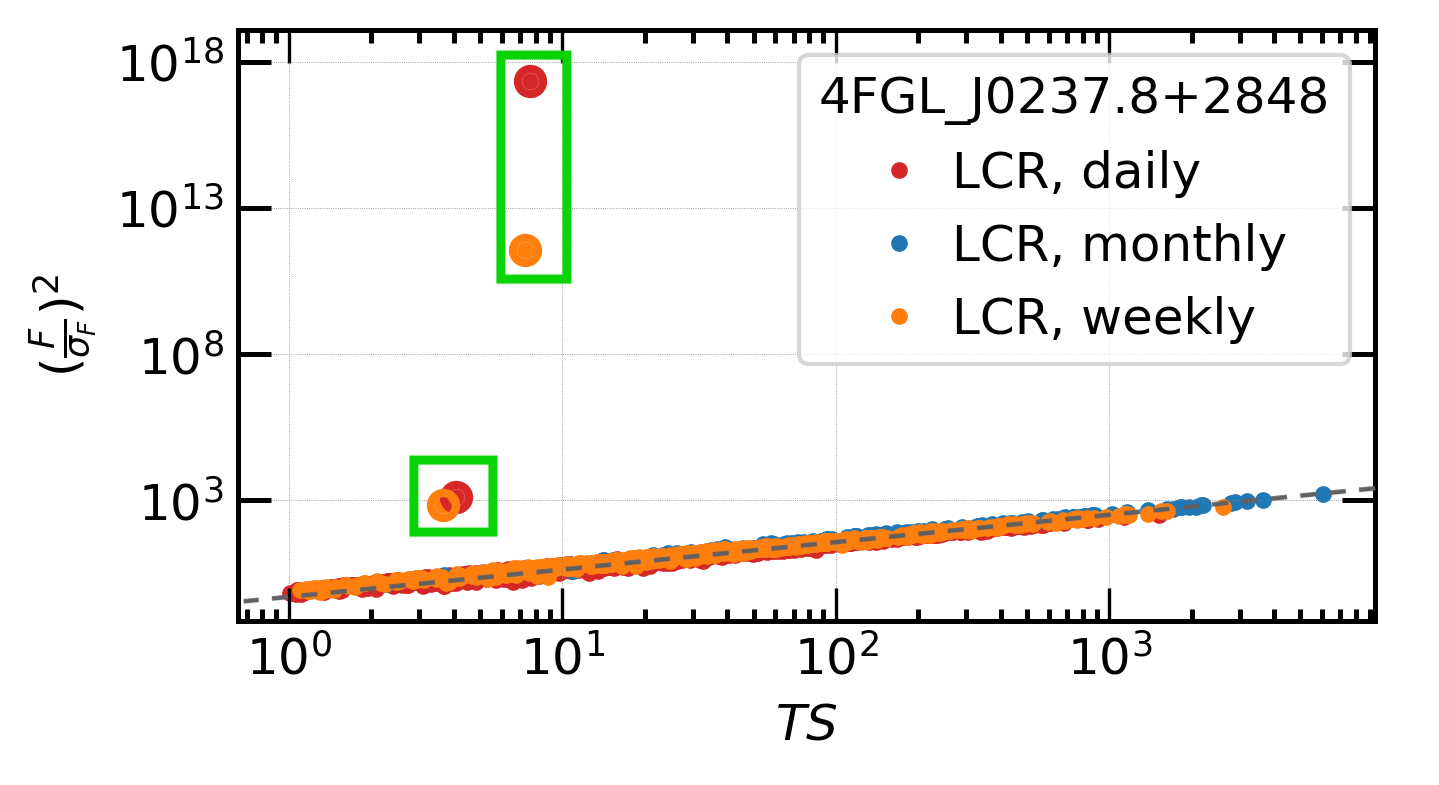}
    \includegraphics[width=0.66\textwidth,angle=0]{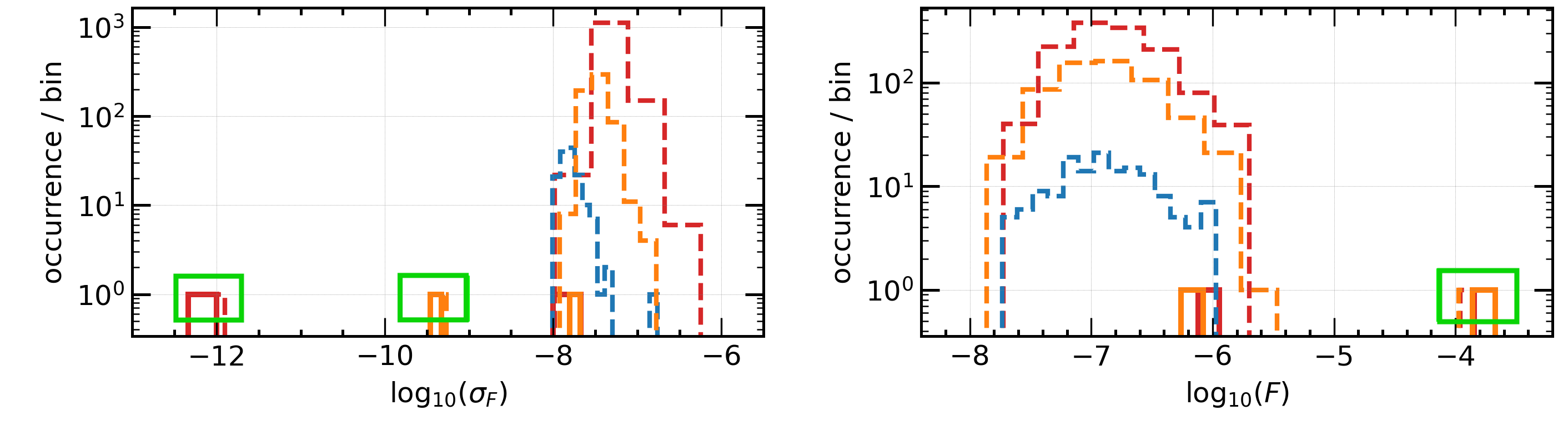}
    \includegraphics[width=0.33\textwidth,angle=0]{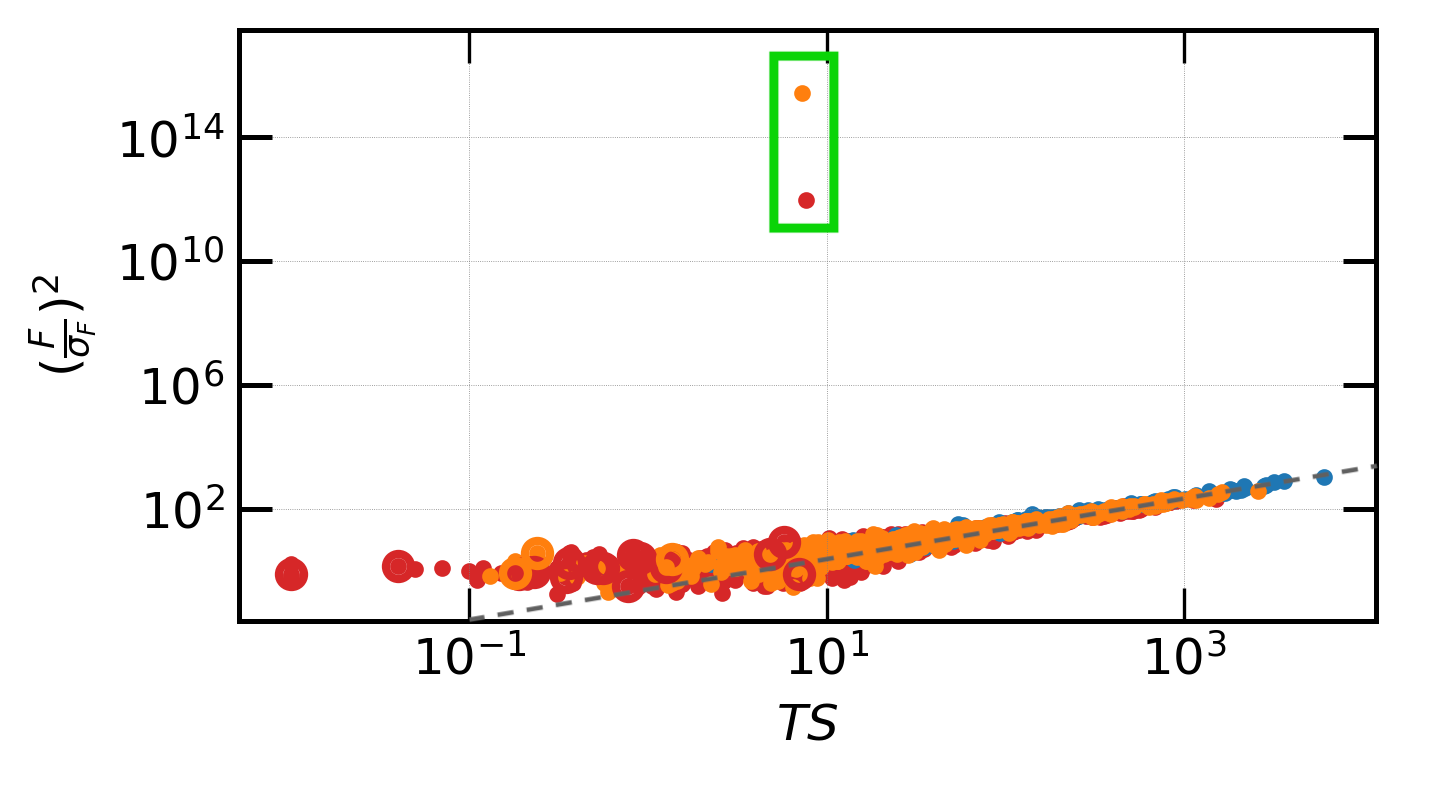}
    \includegraphics[width=0.66\textwidth,angle=0]{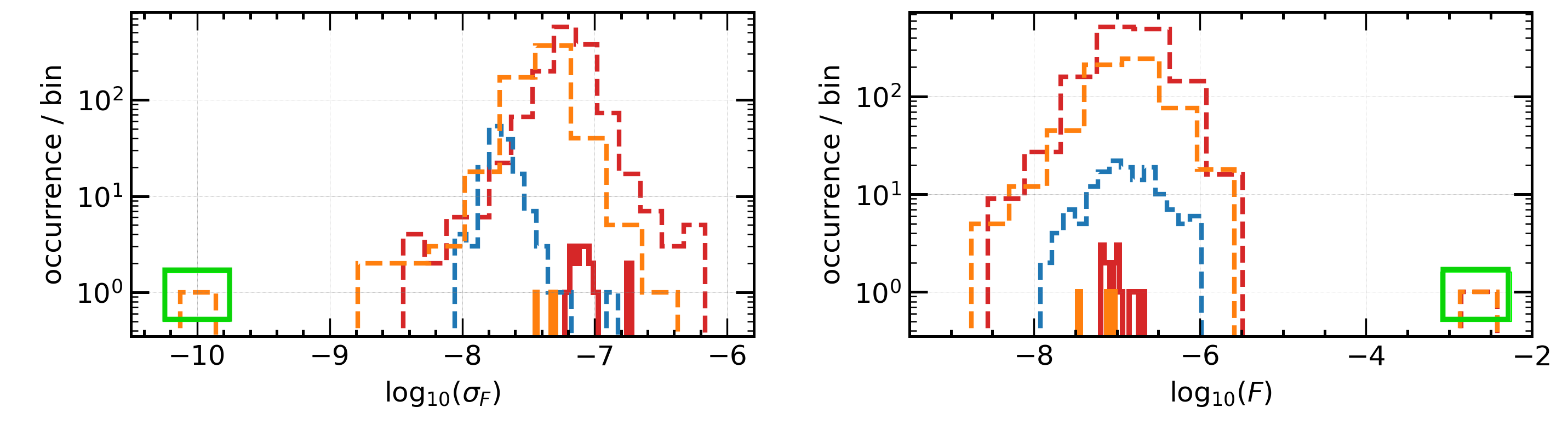}
    \caption{Example of validation plots for the  FSRQ 4C +28.07. The top panels show the case for which the spectral index ($\alpha$ for this source) is fixed. In all the plots we are considering the 0.1--100 GeV energy range. The bottom panels show the case for which the spectral index is free to vary. 
    The ratio of flux to flux uncertainty (left panels) is expected to be approximately proportional to the square root of the TS.} 
    The middle panels show the distributions of flux statistical uncertainties. Flux distributions are shown in the right panels. In this example, distributions are good except for the outliers highlighted within green squares, which should be either further investigated or removed. In the histograms, dashed lines represent results from all time bins. Solid lines represent bins that did not converge. Notice that outliers are not necessarily the results from analyses that did not converge. 
    \label{fig:valid}
    \end{center}
    \end{figure*} 
    
\begin{enumerate}

    
    \item The LCR provides fit results from likelihood analyses that both did and did not converge. However, it is important that the end user is aware that results from analyses that did not converge should be considered suspect and not be used for higher-level analyses (e.g., multi-frequency cross-correlation, power spectral density, or studies of variability). The convergence status for a particular time bin is recorded in the \texttt{MINUIT Return Code} parameter and non-convergent analyses are hidden by default, but are optionally accessible to the user. A review of the current version of the LCR fit results for the first fourteen years of mission data shows that the analyses for 0.7\% of time bins, for all sources, did not converge when the spectral indices were held fixed, and that $\sim$35\% did not converge when the spectral indices were free to vary.
    
    \item From its definition, equation \ref{eq:ts}, the TS is a manifestly positive quantity. However, negative TS values can sometimes be obtained when the parameters reach the limits of their allowed intervals without having maximized the likelihood profile. Fit results obtained from intervals that resulted in negative TS values should be considered suspect and not used in higher-level analyses. The current version of the LCR data products for the first fourteen years of the mission for all LCR sources have only a few bins with negative TS results, per cadence, for both the fixed and free fits.
    
    \item While time intervals containing gamma-ray bursts (GRBs) and solar ﬂares have been removed from the LAT data prior to the likelihood analyses, possible contamination by the proximity of the quiescent Sun has not been accounted for, nor have those time ranges been excluded. The angular separation of the Sun from the target source is provided for each time bin. A total of 175 GRBs and 266 solar flare time intervals were excluded from the data prior to performing the maximum likelihood analysis for the LCR first fourteen years of light curve data. This affects less than about 0.01\% of the time bins for all the LCR sources per cadence.
    
    \item Because the LCR analyses are made available in real-time (new analysis results are generally made available within 24 hours of being processed), the results are not validated by the LAT Collaboration prior to release. Users are encouraged to perform sanity checks by examining the ratio of flux to flux uncertainty vs. the square root of the TS, and the distributions of fit results, e.g., ﬂux, ﬂux uncertainties, spectral indices mean values (photon index $\Gamma$ for PL, $\alpha$ for LP, and $\gamma_1$ for the PLEC model) and their uncertainties. Some examples of these are shown in Fig.~\ref{fig:valid} for the specific case of the  FSRQ 4C +28.07 (4FGL J0237.8+2848). The ﬂux over the ﬂux uncertainty ratio is expected to be approximately proportional to the square root of the TS. Any outliers should be either further investigated or removed before using the data for higher-level analyses, as should any extreme outliers from the data distributions. 

    \item The free-spectral-index light curves provided by the LCR were produced using a model of the pertinent region of the sky for which only the spectral index of the target source is set free, and those of all the other sources were fixed to the 4FGL-DR2 catalog values. Therefore, contamination induced by possible changes in the spectral indices of the sources surrounding the target are not taken into account. For instance, sources undergoing bright flares have been seen to also experience dramatic changes in their spectral indices, including changes in the curvature, at the same time (e.g., harder-when-brighter behavior). Therefore, bright, variable sources in the ROI can induce this type of contamination and must be considered  when a target source is in close proximity to any bright variable sources.
    
    \item Some erroneously high flux values during periods of zero or low exposure, often associated to a very small (or zero) error, have been found for several sources (see Section \ref{subsec:exposure}). We recommend checking the exposure value indicated for each time bin and source ROI in the provided tables. If the light curve has fluxes with very small fractional uncertainties, rather than flux upper limit, in the time bins with zero or low exposure, the reported fluxes and uncertainties should be considered unreliable and those time bins should be excluded from consideration. Time bins with zero error on the flux estimations are automatically hidden but are optionally available to the user.

    \item  A bug in the {\tt make4FGLxml.py} tool that the LCR uses to generate the ROI models was recently identified, in which the reported extension of the extended sources with {\tt RadialGauss} spatial profile is the 68\% C.L. value instead of the sigma value used in the XML model. Thirteen 4FGL-DR2 sources have this spatial profile, namely, Crab IC, IC 443, Monoceros, HESS J1303-631, FHES J1501.0$-$6310, FHES J1626.9$-$2431, FHES J1723.5$-$0501, HESS J1825$-$137, W~41, Cygnus Cocoon, FHES J2129.9+5833, FHES J2208.4+6443 and FHES J2304.0+5406. None of these sources is deemed variable in the 4FGL-DR2 catalog. However, they are present in the ROIs of 101 LCR sources, which might result in a systematic bias in the target flux values across the whole light curve. 
    Correcting this issue will require reprocessing the data for these sources, which is already underway. Meanwhile users should be mindful of this issue when considering sources in the vicinities of these extended sources.
    
    \item Particular care should be taken when using the light curve for the synchrotron component of the Crab. The Crab has different components treated as separate entries in the 4FGL-DR2 catalog: the extended emission, the inverse Compton emission from the PWN and the synchrotron emission from the pulsar. The LCR analysis generates only the light curve for the synchrotron (variable) emission, while keeping the parameters of the other two components frozen in the fit. This could result in some contamination of the synchrotron component light curve deriving from any unaccounted-for variability of the other Crab components. On the other hand, the LCR approach overrides the apparent (not real) variability reported for the pulsar in 4FGL-DR2 (and all other versions of the 4FGL catalog).
    
\end{enumerate}

\section{Prospects and conclusions}
\label{sec:conclusion}

The development of the Fermi-LAT LCR was motivated by the need for a coherent collection of light curves of variable gamma-ray sources observed by the {\sl Fermi}-LAT in support of the time-domain and multi-messenger communities. By continuously reporting the flux evolution and transition to high-flux states for many variable sources, the LCR is a valuable resource for triggering observations of other observatories. Furthermore, the LCR can be used to validate the study of variable activity in neighboring faint sources, helping to identify potential contamination from flaring activity of a bright source.  
In this manuscript we described the automated analysis pipeline which will continuously update the repository with new data as soon as new observations by the LAT become available. We invite the community to use the LCR data products, and report any issues or suggestions to the LCR contacts at the Fermi Science Support Center. A brief guide for navigating the LCR to navigate the web site is provided in Appendix.

In the future, the LCR source list and the resulting light curve data will be updated with every main release of the \textit{Fermi}-LAT source catalog, with the next LCR version planned to be reprocessed entirely and released alongside the 5FGL catalog. In each new release, the LCR will continue following the latest analysis recommendations of the FSSC. Due to the computational expense of re-analyzing the full-mission light curves for the entire LCR sample, the LCR sample will not be updated for each catalog sub-release (e.g., the newly available 4FGL-DR3 \citep{4fglDR3} and the upcoming 4FGL-DR4), or for model and software updates (e.g. related to diffuse background components, PSF and the \texttt{Fermitools}), unless we think that their impact on the light curve is significant. Currently, the LCR does not provide permanent identifiers that allow distinguishing between different versions due to data reprocessing. This will be implemented in a future version of the repository. 

In conclusion, in this era of large surveys, the {\sl Fermi}-LAT is the only high-energy gamma-ray observatory to continuously monitor variable sources, providing the all-sky coverage to identify gamma-ray counterparts to transient events at other wavelengths. We expect that the LCR will greatly enhance the usefulness of LAT data to the time-domain, multi-messenger and multi-wavelength communities.


\section*{Acknowledgments}
MN and JV acknowledge that the material is based upon work supported by NASA under award number 80GSFC21M0002. DK and MN acknowledge support to this work from NASA Fermi GI Program under grant number 80NSSC23K0242. AB is supported by the NASA Postdoctoral Program at NASA Goddard Space Flight Center, administered by Oak Ridge Associated Universities.
INFN and ASI personnel performed in part under ASI-INFN Agreements No. 2021-43-HH.0. Work at NRL is supported by NASA.
The \textit{Fermi} LAT Collaboration acknowledges generous ongoing support
from a number of agencies and institutes that have supported both the
development and the operation of the LAT as well as scientific data analysis.
These include the National Aeronautics and Space Administration and the
Department of Energy in the United States, the Commissariat \`a l'Energie Atomique
and the Centre National de la Recherche Scientifique / Institut National de Physique
Nucl\'eaire et de Physique des Particules in France, the Agenzia Spaziale Italiana
and the Istituto Nazionale di Fisica Nucleare in Italy, the Ministry of Education,
Culture, Sports, Science and Technology (MEXT), High Energy Accelerator Research
Organization (KEK) and Japan Aerospace Exploration Agency (JAXA) in Japan, and
the K.~A.~Wallenberg Foundation, the Swedish Research Council and the
Swedish National Space Board in Sweden.
 
Additional support for science analysis during the operations phase is gratefully
acknowledged from the Istituto Nazionale di Astrofisica in Italy and the Centre
National d'\'Etudes Spatiales in France. This work performed in part under DOE
Contract DE-AC02-76SF00515.


\bibliography{LCRref}{} 
\bibliographystyle{aasjournal}

\appendix
\section{Quick User Guide}
\label{sec:guide}

In this appendix we provide a comprehensive list of the main features of the LCR website\footnote{\url{https://fermi.gsfc.nasa.gov/ssc/data/access/lat/LightCurveRepository/}}.
\begin{itemize}
    \item The main page of the website features an interactive \textit{Catalog Map}
    plotting the positions of all 4FGL-DR2 sources. Optionally, additional data may be overlaid on the map , e.g., real time Sun or Moon position, positions of IceCube neutrino alerts, and GRB error circles as reported in the Second LAT GRB catalog \citep{Ajello2019}. The 1525 LCR sources are highlighted in dark gray, while by default the non-variable 4FGL-DR2 sources are marked in light gray. Hovering over any source displays a \textit{tooltip} box showing its name and key characteristics as well as linking to its 4FGL light curve and spectrum, related FAVA entry, and LCR light curve if applicable\footnote{This option can be disabled by unchecking the \textit{Source info} option from the Map Options menu.}. 
    Below the map, a table is shown listing the 4FGL sources and important parameters. Clicking the name of a source included in the LCR opens a separate page dedicated to that source. 

    \item A \textit{Map Options} menu provides numerous options related to the display of the Catalog Map. Options are provided to change the coordinate system and celestial projection. The marker label and color can be changed, as can the meaning of its size to indicate the variability index, average significance, or time-resolved significance in 3-day bins. 
    
    \item A \textit{Catalog Search} toolbox allows the user to search for a specific source by name or Right Ascension and Declination. The results of the search are highlighted in the Catalog Map. Clicking on the linked name of the target source opens a dedicated page for the source in a new tab. 
    
    \item A \textit{Data Overlays} toolbox allows the user to visualize a number of additional catalog overlays in the sky map. These catalogs include: the \textit{Fermi}-LAT Gamma-ray Burst Catalog \citep[2FLGC; ][]{Ajello2019}, the IceCube Neutrino Alerts\footnote{\url{https://gcn.gsfc.nasa.gov/amon.html}}, and the FAVA Flare Catalog \citep[2FAV;][]{Abdollahi2017}. Activating any of the catalogs will also add the related table under the map.

    \item The dedicated page for each source shows an interactive light curve displaying the detections and upper limits for that source. Options are provided to show the 3 day, 7 day (1 week), or 30 day (1 month) cadence light curve. By default only the significant flux points (with TS$\geq$4) are shown, while upper limits are shown for less-significant time bins. However the user can choose to change the minimum detection significance through the drop-down menu, by selecting among the available options (TS$_{min}$=4, 3, 2, 1). The \textit{Spectral Fitting} option allows the user to choose to visualize either the best-fit values obtained with the fixed spectral index fit or the ones resulting from the fit with spectral index free to vary. A table listing the main characteristics of the selected source is provided. Additional information about the fit convergence, fit tolerance and detection ratio is reported for diagnostics purposes below the light curve plot. 

    \item A \textit{Data Download} toolbox on each dedicated source page provides all data for that source for download. The data are provided in CSV and JSON formats\footnote{A description of the file formats can be found at \url{https://fermi.gsfc.nasa.gov/ssc/data/access/lat/LightCurveRepository/table_description.html}}. All data points are provided, potentially including unconstrained and possibly TS$<0$ data points, or from analyses that did not converge. Guidelines for cleaning the data before use in analysis are given in Sec.~\ref{sec:caveats}.

    \item Finally, the LCR contains a \textit{Usage Notes} page which reviews the data analysis and modeling details, fitting strategy, and caveats for usage (see Sec.~\ref{sec:caveats}).
\end{itemize}



\end{document}